\documentclass[aip,rsi,reprint,graphicx]{revtex4-1}
\usepackage{color}
\usepackage{amsmath}
\usepackage{graphicx}
\usepackage{multirow}
\usepackage{amsfonts}
\usepackage{amssymb}
\usepackage{amscd}

\begin{document}

\title{Fully integrated 3.2 Gbps quantum random number generator with real-time extraction}

\author{Xiao-Guang Zhang}
\affiliation{Hefei National Laboratory for Physical Sciences at the Microscale and Department
of Modern Physics, University of Science and Technology of China, Hefei, Anhui 230026, China}
\affiliation{CAS Center for Excellence and Synergetic Innovation Center in Quantum Information
and Quantum Physics, University of Science and Technology of China, Hefei, Anhui 230026, China}

\author{You-Qi Nie}
\affiliation{Hefei National Laboratory for Physical Sciences at the Microscale and Department
of Modern Physics, University of Science and Technology of China, Hefei, Anhui 230026, China}
\affiliation{CAS Center for Excellence and Synergetic Innovation Center in Quantum Information
and Quantum Physics, University of Science and Technology of China, Hefei, Anhui 230026, China}

\author{Hongyi Zhou}
\affiliation{Center for Quantum Information, Institute for Interdisciplinary Information Sciences,
Tsinghua University, Beijing 100084, China}

\author{Hao Liang}
\affiliation{Hefei National Laboratory for Physical Sciences at the Microscale and Department
of Modern Physics, University of Science and Technology of China, Hefei, Anhui 230026, China}
\affiliation{CAS Center for Excellence and Synergetic Innovation Center in Quantum Information
and Quantum Physics, University of Science and Technology of China, Hefei, Anhui 230026, China}

\author{Xiongfeng Ma}
\affiliation{Center for Quantum Information, Institute for Interdisciplinary Information Sciences,
Tsinghua University, Beijing 100084, China}

\author{Jun Zhang}
\email{zhangjun@ustc.edu.cn}
\affiliation{Hefei National Laboratory for Physical Sciences at the Microscale and Department
of Modern Physics, University of Science and Technology of China, Hefei, Anhui 230026, China}
\affiliation{CAS Center for Excellence and Synergetic Innovation Center in Quantum Information
and Quantum Physics, University of Science and Technology of China, Hefei, Anhui 230026, China}

\author{Jian-Wei Pan}
\affiliation{Hefei National Laboratory for Physical Sciences at the Microscale and Department
of Modern Physics, University of Science and Technology of China, Hefei, Anhui 230026, China}
\affiliation{CAS Center for Excellence and Synergetic Innovation Center in Quantum Information
and Quantum Physics, University of Science and Technology of China, Hefei, Anhui 230026, China}

\date{\today}

\begin{abstract}
We present a real-time and fully integrated quantum random number generator (QRNG) by measuring laser phase fluctuations. The QRNG scheme based on laser phase fluctuations is featured for its capability of generating ultra high-speed random numbers. However, the speed bottleneck of a practical QRNG lies on the limited speed of randomness extraction. To close the gap between the fast randomness generation and the slow post-processing, we propose a pipeline extraction algorithm based on Toeplitz matrix hashing and implement it in a high-speed field-programmable gate array. Further, all the QRNG components are integrated into a module, including a compact and actively stabilized interferometer, high-speed data acquisition, and real-time data post-processing and transmission. The final generation rate of the QRNG module with real-time extraction can reach 3.2 Gbps.

\end{abstract}


\maketitle 

Random numbers are required in many applications such as numerical simulations, cryptography and even lotteries.
Quantum random number generators (QRNGs)~\cite{MaQRNG}, exploiting the basic principles of quantum physics, can produce true random numbers which are unpredictable, irreproducible, and unbiased.
So far, various QRNG schemes have been proposed and experimentally demonstrated~\cite{Rarity94,Stefanov00,Jennewein00,Ma05,Dynes08,Wayne10,Furst10,Qi10,
Gabriel10,Wahl11,Symul11,Jofre11,Bustard11,Jian11,Marandi12,Xu12,Nie14,Yuan14,Sanguinetti14,Yan14,Nie15}.
These QRNG schemes can be simply sorted into three categories. The first one is beam splitter scheme by measuring the path selection when a single photon passes through a beam splitter\cite{Rarity94,Stefanov00,Jennewein00}. In such scheme, one bit at most is generated per photon detection and QRNG speed 
is limited by the count rates of single-photon detectors.
The second one is time measurement scheme by measuring and digitizing photon arrival times~\cite{Ma05,Dynes08,Wayne10,Furst10,Wahl11,Nie14}, and QRNG speed in this scheme reaches roughly 100 Mbps~\cite{Wayne10,Wahl11,Nie14}.
The third one is quantum fluctuation scheme by measuring vacuum states~\cite{Gabriel10,Symul11,Jofre11,Bustard11,Marandi12} or measuring laser phase fluctuations~\cite{Qi10,Xu12,Yuan14,Nie15}, in which classical photodetectors are used instead of single-photon detectors and thus the generation rate can be greatly increased up to Gbps.

In the scheme of laser phase fluctuations~\cite{Qi10,Xu12,Nie15}, the randomness originates from laser spontaneous emission. Given a laser operated around its threshold level,
the contribution ratio between spontaneous and stimulated emissions can be pretty high so that quantum noise dominates the phase fluctuations.
Further, phase fluctuations can be converted into intensity fluctuations using an interferometer, which can be measured by a fast photodetector.
The output signals of the photodetector are digitized to generate raw random data.
To generate final random numbers the min-entropy of the raw random data is evaluated and the bias is removed with randomness extraction.
Owing to the fast photodetector and the high-speed digitizer as well, the final random bit rate can be extremely high.
For instance, a record QRNG speed of 68 Gbps has been recently reported based on the scheme of laser phase fluctuations~\cite{Nie15}.

However, we remark that such high-speed QRNGs have their limitations for practical use~\cite{Ma15}.
In the previous experiments~\cite{Qi10,Xu12,Nie15}, the photodetector outputs were digitized by oscilloscopes,
and the sampled raw random numbers were temporarily stored in the limited memories of the oscilloscopes, which were then post-processed offline.
In addition, compact and integrated QRNG modules are highly required for practical applications.

To close the gap between experimental demonstration and practical use,
here we report a real-time, fully integrated and standalone instrument of QRNG with a generation rate of 3.2 Gbps based on the scheme of laser phase fluctuations. The
random bitstream is transmitted via a small form-factor pluggable (SFP) optical transceiver. This real-time bit rate is higher than the fastest commercially available QRNG product with over 20 times.


\begin{figure}
\centering
\includegraphics[width=4.9 cm]{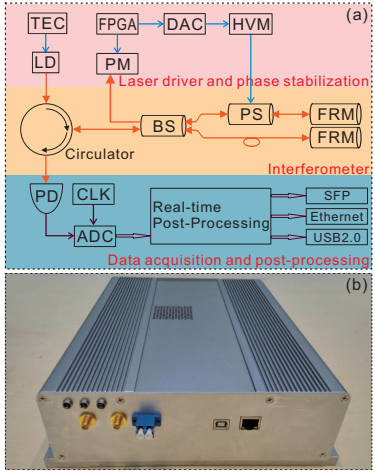}
\caption{
Design diagram (a) and photo (b) of the QRNG module.
}
\label{fig1}%
\end{figure}

The design diagram of the QRNG module is described in Fig.~\ref{fig1}(a).
The key components of the QRNG module
include a stable interferometer that is assembled inside a compact box and two printed circuit boards (PCBs). One PCB includes a laser diode driver with temperature control and a circuit for phase stabilization, whilst the other PCB is designed for raw data acquisition, real-time post-processing and final random data transmission. All the components are integrated into a metal box with a size of 304 mm$\times$250 mm$\times$78 mm as shown in Fig.~\ref{fig1}(b).

A 1550 nm laser diode (LD) is driven by constant current that is slightly above its threshold, and a thermoelectric cooler (TEC) is employed to stabilize the temperature of LD. The emitted continuous wavelength photons enter an unbalanced interferometer
via an optical circulator. One port of the circulator is connected with a 50/50 beam splitter (BS), whose output ports are connected with two Faraday rotator mirrors (FRMs) to construct a polarization-insensitive Michelson interferometer. A phase shifter (PS) is inserted in one arm of the interferometer, and the time difference between the two arms is around 0.8 ns that is much smaller than the coherence time of LD. One output port of the interferometer is detected by a 9.5 GHz InGaAs photodetector (PD) after passing through the circulator, whilst the other output port is monitored by a power meter (PM). The PM is implemented using another PD to measure the optical power and an analog-to-digital converter (ADC) to digitize the measured signal. A field-programmable gate array (FPGA) reads out the PM data, and sends feedback data to a digital-to-analog converter (DAC) to form a voltage signal after the computation by a proportional-integral-derivative (PID) algorithm. The voltage signal regulates the output of a high-voltage module (HVM), which results in automatic adjustments to the PS.
Due to real-time and fast responses of the PID algorithm, the interferometer can be highly stable, which guarantees continuous operations of the QRNG module.

To obtain raw random data, the voltage output of the PD is amplified and then digitized by an 8-bit ADC (TI ADC083000) with a clock (CLK) of 1 GSa/s,
and the sampled data are then fed into a high-speed FPGA (Xilinx Virtex-6).
A pipeline post-processing algorithm based on Toeplitz hashing matrix is implemented in this FPGA for real-time randomness extraction.
The extracted random numbers are transmitted in real-time via the interface of SFP with 3.2 Gbps.
For the applications requiring lower bit rates, either the Gbps Ethernet port or the universal serial bus (USB) 2.0 port can be used.

To quantify the randomness of the raw random data
min-entropy evaluation is applied. The min-entropy is defined as
$H_{min}(X)=-\log_2(\max_{x\in\{0, 1\}^N} P_r[X=x])$,
which can be exploited to quantify the extraction ratio between the raw random bits and the final random bits given a probability distribution of $\{0,1\}^N$ ~\cite{Xu12,Ma13}.
Detailed modeling and analysis of min-entropy for the scheme of laser phase fluctuations can be found in the literatures~\cite{Qi10,Xu12,Nie15,Ma15}.
The key parameter for the evaluation is the ratio of the quantum phase fluctuations to classical noise ($\gamma$).
For our QRNG module, we follow the same experimental approach to measure $\gamma$ by tuning the laser power~\cite{Xu12,Nie15} and repeat this measurement many times. The value of optimal $\gamma$ is stable, and a typical measured value is 6.87.
The variance of quantum phase fluctuations can be calculated~\cite{Qi10,Xu12,Nie15,Ma15} by
$\sigma_q^2=\frac{\gamma}{\gamma+1}\langle V(t)^2\rangle$,
where $\sigma_q$ is the standard deviation of the raw random data, and $V(t)$ is the voltage output amplitude of PD. In our QRNG module,
the measured intensity variance is around 8311 $mV^2$, therefore, $\sigma_q$ is 85.2 mV. Then, one can calculate the maximum probability in the whole distribution is 0.011 since the quantum signal follows a Gaussian distribution~\cite{Ma13}. $H_{min}(X)$ is thus calculated as 6.5 bits per sample or 0.8 bits per raw bit, which means that 6.5 random bits can be generated from each sample.

\begin{figure}
\centering
\includegraphics[width=4.9 cm]{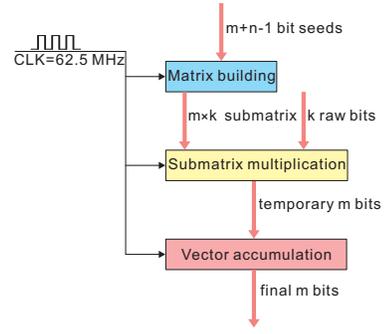}
\caption{FPGA implementation of Teoplitz hashing randomness extraction.
\label{fig2}}%
\end{figure}

Since the capability limit of real-time post-processing in FPGA is 5 Gbps, 3 bits are abandoned for each sample including the least significant bit and the left two bits. Considering the worst case
there are still 3.5 bits can be extracted from each sample, which corresponds to a new min-entropy of 0.7 bits per raw bit.

After the min-entropy evaluation, a Toeplitz hashing extractor is applied to distill the raw random data,
and the rigorous discussion about Teoplitz hashing extractor can be found in the reference~\cite{Ma13}.
Given a binary Toeplitz matrix with a size of $m\times n$, $m$ final random bits are extracted by multiplying the matrix and $n$ raw bits.
The Toeplitz matrix can be constructed by a sequence of $m+n-1$ random bits called matrix building seeds, due to its characteristic
that all the elements of each descending diagonal from left to right are the same.

For our implementation of Toeplitz hashing extraction, we choose $m=1024$ and $n=1520$ so that the extraction ratio is $m/n=0.67$. According to the Leftover Hash Lemma~\cite{Impagliazzo89},
$m=nH_{min}(X)-2\log_2(\frac{1}{\epsilon})$,
one can calculate the information theoretic security bound $\epsilon=2^{-20}$,
that is, the statistical distance between the extracted random sequence and the uniform sequence is bounded by $\epsilon=2^{-20}$.
The extraction efficiency is ${m}/{[nH_{min}(X)]}=0.96$.


\begin{figure}[b]
\centering
\includegraphics[width=8 cm]{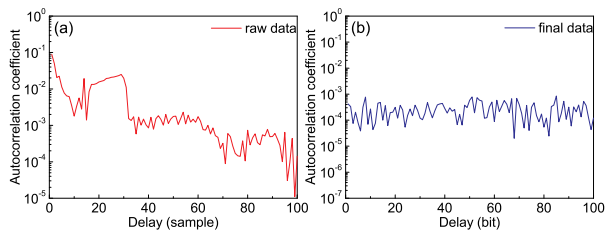}
\caption{Autocorrelation analysis of $10^7$ raw samples (a) and $10^7$ extracted random bits (b). The autocorrelation existing in the raw data is completely eliminated by the Toeplitz hashing extraction.
\label{fig3}}%
\end{figure}

In the QRNG module, the Toeplitz hashing randomness extraction is implemented in FPGA. By taking advantages of concurrent computation capability in FPGA, the real-time post-processing speed can be extremely improved compared with the software implementation in computer~\cite{Nie15}. Due to the resource limit in FPGA, it is impossible to directly compute such a large matrix rapidly. Therefore, we propose a concurrent pipeline algorithm to achieve the real-time extraction.
In particular, in order to implement real-time randomness extraction, we design three modules including matrix building, submatrix multiplication and vector accumulation in FPGA, as shown in Fig.~\ref{fig2}. These modules are operated in a pipeline mode with a synchronized clock of 62.5 MHz. In the matrix building module, the $m+n-1 (=2543)$ bit seeds are used to construct the complete Toeplitz matrix. These seeds can be refreshed periodically. The whole Toeplitz matrix is divided to $n/k$ submatrices. Considering the resource consumption in FPGA, the value of $k$ is selected to be 80. In each clock period a submatrix with a size of $m\times k$ and $k$ raw bits are multiplied to output a temporary column vector in the submatrix multiplication module. After $n/k$ clocks, all the calculations for $n$ raw bits are finished. The $n/k$ temporary column vectors are then added in the vector accumulation module and $m$ final random bits are generated.
We note that binary multiplications and binary additions can be realized by bitwise AND and bitwise XOR operations, respectively. With such a configuration, real-time Toeplitz hashing randomness extraction is achieved in the FPGA, and the generation rate limit of final random bits can reach 5 Gbps $\times 1024/1520 \sim$ 3.36 Gbps.
The final generation rate in FPGA can be tuned to match the interfaces for the real-time transmission.
With a SFP optical transceiver, the real-time rate of the final random numbers reaches 3.2 Gbps.
Apart from the SFP, optional interfaces including Gbps Ethernet port and USB 2.0 port are also designed for the applications requiring lower random bit rates, whose average speeds are tested as 968.7 Mbps and 259.5 Mbps, respectively.


To test the randomness of the final data, we perform an autocorrelation comparison between the raw data and the extracted random data, as shown in Fig.~\ref{fig3}. The relatively large autocorrelation existing in the raw data is mainly due to the short sampling interval~\cite{Ma15}.
For each sample, the interferometer output can be regarded as an interference from two certain points in the same beam with a time difference of $\tau$, where $\tau$ stands for the time delay between the two arms of the interferometer. The two points form a wave interval. When the sampling rate is high enough, one interval overlaps with adjacent intervals, which results in a large autocorrelation.
Fig.~\ref{fig3} shows that after post-processing this autocorrelation can be significantly reduced.
For the randomness quantification, the theory given in the reference~\cite{Ma15} can be applied, which aims at a high sampling rate situation.
Finally, the standard NIST statistical tests is applied to test the randomness of final data. Typically, 10 final random data files with each file size of 1 Gbits are tested, and all the files can well pass the test items. We note that given the items that produce multiple outcomes the $p$-values are processed by a Kolmogorov-Smirnov uniformity test and the proportions are averaged.



In summary, we have developed a real-time and fully integrated QRNG module
based on the scheme of laser phase fluctuations.
We propose and implement a pipeline post-processing algorithm based on Toeplitz hashing randomness extraction in FPGA, which can extremely increase the real-time generation rate of final random numbers up to 3.2 Gbps.

This work has been financially supported by the National Basic Research Program of China Grant No.~2013CB336800, the National Natural Science Foundation of China Grant No.~61275121, and the Chinese Academy of Sciences. X.-G. Zhang and Y.-Q. Nie contributed equally to this work.

\end{document}